\documentstyle[prl,aps,preprint]{revtex}

\newcommand{\be}{\begin{equation}}
\newcommand{\ee}{\end{equation}}
\newcommand{\ba}{\begin{eqnarray}}
\newcommand{\ea}{\end{eqnarray}}
\newcommand{\bec}{\begin{center}}
\newcommand{\eec}{\end{center}}

\begin{document}
\draft

\widetext

\title{Supersymmetric quantum cosmology \\
for Bianchi class A models}

\author{
Alfredo Mac\'{\i}as$^\diamond$\thanks{E-mail: amac@xanum.uam.mx}, 
Eckehard W. Mielke$^\diamond$\thanks{E-mail: ekke@xanum.uam.mx},
and Jos\'e Socorro$^{\$}$\thanks{E-mail: socorro@ifug4.ugto.mx}\\
$^{\diamond}$ Departamento de F\'{\i}sica,\\
Universidad Aut\'onoma Metropolitana--Iztapalapa,\\
Apartado Postal 55-534, C.P. 09340, M\'exico, D.F., MEXICO\\
$^{\$}$ Instituto de F\'{\i}sica de la Universidad de Guanajuato,\\
Apartado Postal E-143, C.P. 37150, Le\'on, Guanajuato, MEXICO
}

\date{\today}

\maketitle

\begin{abstract}
The canonical theory of (${\cal N}=1$) supergravity, with a matrix representation for 
the gravitino covector--spinor, is applied to the Bianchi class A spatially
homogeneous cosmologies. The full Lorentz constraint and its implications for
the wave function of the universe are analyzed in detail. 
We found that in this model no physical states other than the
trivial ``rest frame" type occur.

\end{abstract}
\vspace{0.5cm}


\narrowtext

{\em Introduction}

Recently we realized the lack of information existing in the literature about 
the implications of the Lorentz constraint on the wave function of the 
universe. In the supersymmetric approach to quantum cosmology,
there exist only vague discussions of how Lorentz invariants should look like 
and how many of them they are. Thus, the problem is usually avoided by 
considering scalar wave functions. These fermionic and bosonic power 
expansions of the wave function are supposed to automaticaly
fulfill the Lorentz condition \cite{g1,ho,gl,gc}.  
We attacked to some extent this problem \cite{mamilo} in full
generality, and showed its meaning on the structure of the wave function of 
the universe, using the Bianchi type IX cosmological models as an example. 
No physical states in the minisuperspace sector of the theory were found. 
Moreover, only when the Lorentz generator trivializes, the well known 
``rest frame" type \cite{mor,som,maoso} wave function solutions arises. 
However, it is interesting to note that the trivial ``rest frame" type 
solutions exist for an arbitrary Lorentz generator. 

The purpose of this paper is to show that our results remain valid for 
{\em all} minisuperspace Bianchi class A models.

{\em Physical States}

As it is well known, the Hamiltonian form of the $({\cal N}=1)$ supergravity 
Lagrangian can be written as \cite{pi78}
\ba
H= N {\cal H}_\bot + N^i {\cal H}_i + \frac{1}{2} \omega_{0 AB}{\cal J}^{AB}
+ {\overline \Psi}_0 {\cal S} 
\label{sgham}\, ,
\ea
where ${\cal H}_\bot$, ${\cal H}_i$ and ${\cal J}^{AB}$ are the usual 
Hamiltonian, diffeomorphism and rotational Lorentz bosonic constraints, 
respectively, and ${\cal S}$ the supersymmetric fermionic constraint. The 
lapse function $N=e_{0}{}^0$, the shift vector $N_i=e_{i}{}^0$, \, 
$\omega_{0AB}$ and ${\overline \Psi}_0$ are the corresponding Lagrange 
multipliers.  
The supergravity generators satisfy the soft algebra discovered by 
Teitelboim \cite{Tei77}. 
Thus, the physical states $\vert\Psi\rangle$ in 
the supersymmetric quantum cosmology theory have to satisfy the conditions
\be
{\cal S} \vert\Psi\rangle=0\, , \qquad {\cal H}_A \vert\Psi\rangle=0\, ,
\qquad {\cal J}_{AB} \vert\Psi\rangle = 0
\label{phst}\, .
\ee 
Note that the constraint ${\cal S} \vert\Psi\rangle=0$ is the ``square root" 
of the Hamiltonian on account of the relation 
$\left\{{\cal S}(x),{\overline {\cal S}}(x^{\prime}) \right\} = \gamma^A\, 
{\cal H}_A\,\delta(x,x^{\prime})$. Since this implies
${\cal H}_A \vert\Psi\rangle=0$, the second condition is redundant. 
Thus, we will focus only on the Lorentz ${\cal J}_{AB}$
and supersymmetric ${\cal S}$ constraints, which are explicitly given as 
follows \cite{pi78}:
\ba
{\cal J}_{AB} &\equiv& p_{A}{}^{\alpha} e_{B \alpha} - p_{B}{}^{\alpha}
e_{A \alpha} - \pi^\alpha \sigma_{AB} \Psi_\alpha\nonumber\\
 &=& 2p_{[A}{}^{\alpha} e_{B] \alpha} + \tau_{AB0}\nonumber\\
 &=& 2p_{[A}{}^{\alpha} e_{B]\alpha} +  {1\over 2} \Psi_{[A}^T \Psi_{B]}
\label{lorentz}\, ,
\ea
with $\pi^\alpha=\frac{i}{2} \varepsilon^{0 \alpha \delta \beta}
{\overline \Psi}_\delta \gamma_5 \gamma_\beta$ the momentum conjugate to the 
gravitino field. In the last step we have  used the fact that \cite{mimamo}
${\overline \Psi}= \Psi^T C=-i \Psi^T \gamma^0$. 
{\em The Cartan relation} \cite{mie86} 
$T_{\mu\nu\lambda}=\tau_{\mu\nu\lambda}=  
(i/4) {\overline \Psi}_{[\mu\vert} \gamma_\lambda \Psi_{\vert\nu]}$ 
relates the torsion to the spin tensor of the Rarita--Schwinger field and is 
used to eliminate the torsion tensor from the theory, leaving the theory only
with first class constraints. 

The generator of supersymmetry reads
\be
{\cal S} = \varepsilon^{0\alpha\beta\delta} \gamma_5 \gamma_\alpha D_\beta
\Psi_\delta  
\label{susyc}\, ,
\ee
where the usual factor ordering is chosen \cite{mor}.

Instead of the gravitino field, it is rather convenient to use 
the densitized local components $\phi_a = e\, e_a{}^\alpha \Psi_\alpha$
as the basic fields commuting with all non--spinor variables, here 
$e={}^{(3)}e= \det(e^\alpha{}_a)$. This variable
was also found to be the natural one for the gravitino field \cite{dks77}.
This choice suggests a matrix realisation of the $\phi_{i A}$ obeying
$\lbrace \phi_{i {\cal A}}, \phi_{j {\cal B}} \rbrace = -{i\over 8} 
 (\gamma_i \gamma_j)_{{\cal A}{\cal B}}$. 
Here ${\cal A}$ and ${\cal B}$ are spinor indices, and the gravitational 
variables appear nowhere.

We will assume the following form for the wave function of the universe
\be
\vert\Psi\rangle=\pmatrix{\Psi_I\cr                                   
            \Psi_{II}\cr
            \Psi_{III}\cr
            \Psi_{IV}\cr}  
\label{wfu}\, .
\ee

{\em Lorentz condition}

It can be easily shown that the bosonic part of (\ref{lorentz}) vanishes 
identically when it is written in one basis, i.e. when the coordinates
are fixed. For the fermionic part of (\ref{lorentz}) we do not expect 
something similar to happen. In fact, for the Bianchi class A models the 
Lorentz generator (\ref{lorentz}) reads:
\be
{\cal J}_{ab} = - {\cal J}_{ba}= {1\over 2} \phi^T_{[a} \phi_{b]} 
={1\over 2} \phi^T_{[a {\cal A}} \phi_{b]}^{\cal A}
\label{lorentz2}\, .   
\ee

Thus, the condition (\ref{phst}) for the physical states can be 
written as
\be
{\cal J}_{AB}\vert\Psi\rangle = \pmatrix{0&~~0&~~0&~0\cr
                         0&~0&~{\cal J}_{12}&~{\cal J}_{13}\cr
                         0&~-{\cal J}_{12}&~~0&~{\cal J}_{23}\cr
                         0&~-{\cal J}_{13}&~- {\cal J}_{23}&~~0\cr}   
~~ \pmatrix{\Psi_I\cr                                   
            \Psi_{II}\cr
            \Psi_{III}\cr
            \Psi_{IV}\cr}= 0
\label{lor}\, .
\ee
In terms of 
${\cal J}_A = \frac{1}{2} \epsilon_{0ABC}{\cal J}^{BC}\quad \Rightarrow \quad
{\cal J}_0=0$,
we can write the conditions (\ref{lor}) as\footnote{These relations are quite 
general, they do not depend on the particular Bianchi model in 
consideration, cf. \cite{Niew81,mie86}}
\be
{\cal J}_3 \Psi_{III} = {\cal J}_2 \Psi_{IV} \, ,\qquad     
{\cal J}_3 \Psi_{II} = {\cal J}_1 \Psi_{IV} \, ,\qquad         
{\cal J}_2 \Psi_{II} = {\cal J}_1 \Psi_{III} 
\label{hsys3}\, ,
\ee
respectively.
It is interesting to note that there is no condition in 
(\ref{hsys3}) involving $\Psi_I$.

On the other hand, Eqs. (\ref{hsys3}) imply that we should 
representate the components ${\cal J}_{ab}$ of the Lorentz generator 
(\ref{lorentz2}) as well as the components $\Psi_i$ of the wave funtion of the
universe (\ref{wfu}) as $4\times 4$ and $4\times 1$ non--singular matrices, 
respectively. If we do so, it is possible to show from (\ref{hsys3})   
that there exists the transformation 
\ba
{\cal J}_1 &=& {\cal J}_3 \left({\cal J}_{2}\right)^{-1} 
{\cal J}_1 \left({\cal J}_{3}\right)^{-1} {\cal J}_2 \, , \nonumber \\  
{\cal J}_2 &=& {\cal J}_1 \left({\cal J}_{3}\right)^{-1} 
{\cal J}_2 \left({\cal J}_{1}\right)^{-1} {\cal J}_3 \, , \nonumber \\     
{\cal J}_3 &=& {\cal J}_1 \left({\cal J}_{2}\right)^{-1} 
{\cal J}_3 \left({\cal J}_{1}\right)^{-1} {\cal J}_2   
\label{pufi}\, .
\ea
There are two possibilities for interpreting the conditions (\ref{pufi}): 
The first one is to regard them as 
an invariance under a certain kind of similarity transformations between the 
different components of the Lorentz generator, implying an underlying 
relation between them.
It would be necessary to find an appropiate matrix representation for the
different components of the gravitino field, i.e. Lorentz operator, which 
should be capable to reflect this invariance.

This can be realized in terms of $\gamma$--matrices via 
the following identification \cite{mamilo}
\be
{\cal J}^1 = - \gamma^1 \gamma^0\, , \qquad {\cal J}^2 = \gamma^1 \gamma^3 \, 
, \qquad {\cal J}^3 =  -\gamma^3 \gamma^0 
\label{repre}\, .
\ee
This representation satisfies also the usual algebra of the angular momentum
generators $\left[{\cal J}^i ,{\cal J}^j\right] = \frac{1}{2}\varepsilon^{ijk}
{\cal J}_k$, of the Lorentz group. Consequently, by solving 
(\ref{repre}) for the components of the gravitino field, we 
find their corresponding matrix representation, namely:
\be
\phi_1 = -i\gamma^3 \, , \qquad \phi_2 =-i \gamma^1 \, , \qquad 
\phi_3 = -i \gamma^0 
\label{papi} \, .  
\ee

The second possibility is to conclude that the Lorentz constraint trivializes,
which means the proportionality of its components
\be
{\cal J}_{3} = \pm {\cal J}_{2} = \pm {\cal J}_{1} 
\label{nosferatu}\, .
\ee 
Consequently, it follows from (\ref{nosferatu}) that
$\phi_1=-\phi_2=\phi_3$
and the components $\phi_{i {\cal A}}$ are thus pure real numbers.
Moreover, the conditions (\ref{nosferatu}) imply for the wave function of the
universe that $\Psi_{II}=\Psi_{III}=\Psi_{IV}$.
As mentioned above, then the Lorentz operator trivializes 
implying that the wave function of the universe should be scalar 
and only two components of it are independent \cite{mamilo}. 
Therefore, in the second interpretation the problem is reduced to consider the
supersymmetric condition for $\Psi_I$ and let say $\Psi_{II}$.
However, this trivialization does not fulfill the constraint algebra 
\cite{Tei77} and should be ruled out.

{\em Bianchi class A models}

The basic field variables in the graviton--gravitino formulation of the $(N=1)$
supergravity are the vierbein $e^A{}_\mu$ and the covector--spinor gravitino
field $\phi_\mu$, which obey the Einstein--Cartan--Rarita--Schwinger system of 
Freedman et. al. \cite{fnf}.  
The general Bianchi type metric can be written in the form 
\be
{ds}^2 ={(N^2 - N^j N_j)}{dt}^2 -{2 N_i~dt~\omega^i}
-{e^{-2{\Omega}(t)}}{e^{{2 \beta}(t)}_{~~ij}}{\omega^i \omega^j}
\label{bianme}\, ,
\ee
where $\Omega (t)$ is a scalar and $\beta_{ij}(t)$ is a $3\times 3$
matrix, and the lapse and shift functions are $N(t)$ and $N_i(t)$ respectively.
The one--forms $\omega^i$ are characteristic for a particular Bianchi 
universe \cite{ryan}. 
They obey the Maurer--Cartan relation 
$d\omega^i = -\frac{1}{2} C^i{}_{jk} \omega^j \wedge \omega^k$,
since $C^i{}_{jk}$ are the structure constants of the particular 
group of motions associated with the Bianchi models. According to the
classification scheme of Ellis and MacCallum \cite{ema}, the structure 
constants are written in the form:
$C^i{}_{jk}= \varepsilon_{jks} m^{si} + \delta^i_k a_j - \delta^i_j a_k$, 
with the matrix ${\bf m}=(m^{ij})$ and the triplet $(a_i)$.
For the class A models $a_i=0$ and \cite{ryan}
\be
{\bf m} = \cases{0 \qquad {\em type}~~ I\cr
{\rm diag}(1,0,0) \qquad {\em type} ~~ II\cr
- {\bf \alpha} \qquad {\em type} ~~ VI_{-1}\cr
{\rm diag}(-1,-1,0) \qquad {\em type} ~~ VII_0\cr
{\rm diag}(-1,1,1) \qquad {\em type} ~~ VIII\cr
\delta_{ij}\qquad {\rm type} ~~ IX\cr}\, ,\qquad {\rm with}\qquad
{\bf \alpha}= \pmatrix{0&1&0\cr
1&0&0\cr
0&0&0\cr}
\ee
We will also take
$\beta_{ij}$ diagonal and use the Misner parametrization \cite{ryan}, 
$\beta_{ij}={\rm diag}(\beta_+ +\sqrt 3 \beta_-, \beta_+ - \sqrt 3 \beta_-, 
-2 \beta_+)$.
Quantum cosmology means the quantization of the homogeneous
models with $\phi_\mu = \phi_\mu (t)$ and the full set of dynamical variables
is $\Omega (t)$, $\beta_{ij}(t)$ and $\Psi_{\mu A_i}(t)$.    

Let us perform a transformation to the natural variables for our
problem, namely:
\be
u^k := \cases{x =  \Omega - \beta_+ - \sqrt{3} \beta_- \, \cr 
y =  \Omega - \beta_+ + \sqrt{3} \beta_- \, \cr
z =  \Omega + 2\beta_+}
\label{natva}\, .
\ee
In terms of the coordinates (\ref{natva}) and in the $\omega^i$ basis 
characteristic of the particular Bianchi model into consideration, the 
vierbein is given by
$e_{0}{}^0= N(t)$\, , $e_{i}{}^0= N_{i}$\, , 
and $e_{j}{}^\alpha= \exp{(u^{(j)})} \delta^\alpha_{j}$, 
where the parenthesis means no summation. The corresponding rotation 
coefficients read \cite{ryan}
$\omega_{ijk}= \frac{1}{2} \exp{(u^{(i)})} C_{ijk}$,
for the Bianchi class A models.
Note that $e= \exp {[-(x+y+z)]}$, so it is easy to find the
local expression for the gravitino field $\phi_a = \exp{[-(x+y+z)]} \Psi_a$.
Using the Majorana representation for the $\gamma$--matrices 
\cite{Niew81,kaku} as well as the matrix representation implied by the 
Lorentz constraint for the components of the gravitino field, we can write 
the supersymmetric generator (\ref{susyc}), for the class A models
as follows
\ba
{\cal S} &=& e^{-1}i\gamma^0 \left\{\gamma^1\left[\left(\gamma^2 \phi_2
+ \gamma^3 \phi_3\right)\partial_x - \frac{1}{2} \exp{(x)} 
(m^2_2+ m^3_3)\left(\gamma^3\phi_2 - \gamma^2 \phi_3\right)\right]\right. 
\nonumber \\
&+&\left.\gamma^2 \left[\left(\gamma^1 \phi_1 + \gamma^3 \phi_3\right)
\partial_y + \frac{1}{2} \exp{(y)} (m^1_1+ m^3_3)\left(\gamma^3 \phi_1
- \gamma^1 \phi_3\right)\right]\right. \nonumber \\
&+& \left. \gamma^3 \left[\left(\gamma^1 \phi_1 + \gamma^2 \phi_2\right)
\partial_z + \frac{1}{2} \exp{(z)} (m^1_1+ m^2_2)\left(\gamma^2 \phi_1
+ \gamma^1 \phi_2\right)\right]\right\} 
\label{ese}\, .
\ea
For the $\gamma$ matrices we use the real Majorana representation 
\cite{Niew81,kaku}, 
in which all the $\gamma$--matrices are purely imaginary and the componentes 
of the gravitino vector spinor are consequently real
with $C = -i\gamma^0$. 

{\em ``Rest frame" type state}

Due to the fact that the Lorentz condition (\ref{lor}) does not constraint
the first component $\Psi_I$ of the wave function of the universe, it is
possible to find the ``rest frame" type solution by choosing
\be
\Psi = \pmatrix{\Psi_I\cr
0\cr
0\cr
0\cr}
\label{trivsol}\, ,
\ee
and then solving the corresponding supersymmetric condition (\ref{ese}). 
It is important to note that the Lorentz generator remains arbitrary, which is
in formal {\em analogy} to rest frame solutions in Wigner's spin and mass 
classification of representations of the Poincar\'e group \cite{kaku}.

In this case the Eqs. (\ref{ese}) can be written as follows 
\be
\left[\left(i\Gamma^1 \partial_1 - \Gamma^2 \frac{a}{2}\exp{(x)}\right)
+\left(i\Gamma^3 \partial_2 + \Gamma^4 \frac{b}{2}\exp{(y)}\right)
+\left(i\Gamma^5 \partial_3 + \Gamma^6 \frac{c}{2}\exp{(z)}\right)\right] 
\Psi_I = 0
\label{trivsus1}\, ,
\ee
where the $\Gamma^i$ are the $4\times 4$ matrices involving the gravitino 
components which appear in (\ref{ese}), $a=m^2_2+ m^3_3$,\, $b=m^1_1+ m^3_3$
\, and \, $c=m^1_1+ m^2_2$.
In order to solve (\ref{trivsus1}), we use the following matrix realization 
for the $\Gamma^i$ matrices in the Majorana representation: 
\be
\Gamma^1 = - \gamma^0 \gamma^1\, , \quad \Gamma^2 = \gamma^0 \, , \quad
\Gamma^3 = - \gamma^0 \gamma^3\, ,\quad
\Gamma^4 = \gamma^0 \gamma^5\, , \quad \Gamma^5 = \gamma^0 \gamma^2\, ,
\quad \Gamma^6 = \gamma^1
\label{dar3}\, .
\ee
Note that $\{\Gamma^A,\Gamma^B\}=0$, with $A \neq B=1,\cdots,6$.

Performing the separation of variables, it is easy to find the solution
of (\ref{trivsus1}), which is given by 
\be
\Psi_{I} =\pmatrix{\Psi_{01}e^{ae^{x}/2}e^{-be^{y}/2}e^{-ce^{z}/2}
                          + e^{-ce^{z}/2}E_{i}(z)l_{1}\cr
                          \Psi_{02}e^{-ae^{x}/2}e^{be^{y}/2}e^{ce^{z}/2}
                          - e^{ce^{z}/2}E_{i}(z)l_{2}\cr
                          \Psi_{03}e^{ae^{x}/2}e^{be^{y}/2}e^{ce^{z}/2}
                          + e^{ce^{z}/2}E_{i}(z)l_{3}\cr
                          \Psi_{04}e^{-ae^{x}/2}e^{-be^{y}/2}e^{-ce^{z}/2}
                          - e^{-ce^{z}/2}E_{i}(z)l_{4}\cr}
\label{trigen}\, ,
\ee
with $l_{j}$ separation constants, and $\Psi_{0j}$ integration 
constants ($j=1,...,4$).  Here we used the exponential
integral function $E_{i}(x)$ \cite{mamilo}.
Note that this unconstrained wave function solution is of solitonic type and 
for arbitrary Lorentz rotation generator. These kind of solutions have been
already obtained in supersymmetric quantum cosmology \cite{mary,cg1,cg2}, and 
also they have been found in the context of the standard Wheeler--DeWitt 
approach \cite{bara,kura,mora,oso}. 

{\em Non--trivial state}

If we look at equations (\ref{ese}), it is easy to conclude that for the
representation induced by satisfying the full non--trivial Lorentz constraint,
the wave function of the universe vanishes for all Bianchi class A models, i.e.
\be
{\cal S}\vert\Psi\rangle=0 \qquad \Rightarrow  \qquad \vert\Psi\rangle=0 
\label{nipedo}\, ,
\ee 
this means that there exist no physical states consistent with the Lorentz 
constraint \cite{cfop}.
{\em Discussion}

In previous papers \cite{mor,som,maoso}, we focused our attention to the
square root property of the supersymmetric constraint, without taking care
of the Lorentz constraint. Since then, many papers
related to the subject have appeared in the literature \cite{som,g1,ho,gl,gc},
none of them consider seriously the issue of the Lorentz constraint 
implications on the wave function of the universe. For the
wave function a particular structure is assumed ad hoc in order to fulfill by 
construction the Lorentz constraint. Thus, our study of the Lorentz
constraint becomes compulsory \cite{mamilo}. 

Here we analyzed in full generality the implications of the Lorentz constraint
on the wave function of the universe, i.e. exactly the meaning 
of ${\cal J} _{AB}\vert \Psi\rangle\cong \tau_{AB0}\vert \Psi\rangle=0$.
The main conclusion of this paper is that for all Bianchi class A 
models, there are no physical states in the supersymmetric approach
to quantum consmology other than the ``rest frame" type, concerning 
only the first component of $\Psi$ which is not constrained by the Lorentz 
condition. The same results are obtained using a singular matrix representation
for the angular momentum generators of the Lorentz group \cite{kaku}.
Our results seem to resolve the apparent discrepancy existing between 
minisuperspace studies and the full supergravity conclusions of Carrol et al. 
in Ref. \cite{cfop}.

{\em acknowledgments}

We would like to thank Michael P. Ryan Jr. for useful discussions and 
literature hints.
This work was partially supported by  CONACyT, grants No. 3544--E9311, No. 
3898P--E9608, and by the joint German--Mexican project KFA--Conacyt 
E130--2924. 
One of us (E.W.M.) acknowledges the support by the short--term
fellowship 961 616 015 6 of the German Academic Exchange Service (DAAD), Bonn.

\end{document}